\documentclass[prd, twocolumn, nofootinbib]{revtex4}

\usepackage{epsfig} 
\usepackage{amsmath}

\newcommand{\beq}{\begin{equation}}
\newcommand{\eeq}{\end{equation}}
\newcommand{\beqa}{\begin{eqnarray}}
\newcommand{\eeqa}{\end{eqnarray}}
\newcommand{\om}{\Omega_m}
\newcommand{\omw}{\Omega_w}
\newcommand{\omwmx}{\Omega_{w,{\rm max}}}

\newcommand{\as}{a_\star}

\newcommand{\op}{\Omega_w}

\newcommand{\ls}{\mathrel{\raise0.27ex\hbox{$<$}\kern-0.70em \lower0.71ex\hbox{{
$\scriptstyle \sim$}}}}

\begin{document} 

\title{Calibrating Dark Energy} 
\author{Roland de Putter \& Eric V.\ Linder} 
\affiliation{Lawrence Berkeley National Laboratory \& University 
of California, Berkeley, CA 94720, USA} 
\date{\today}

\begin{abstract} 
Exploring the diversity of dark energy dynamics, we discover a 
calibration relation, a uniform stretching of the amplitude of the 
equation of state time variation with scale factor.  This defines 
homogeneous families of dark energy physics.  The calibration factor 
has a close relation to the standard time variation parameter $w_a$, 
and we show that the new, calibrated $w_a$ describes observables, i.e.\ 
distance and Hubble parameter as a function of redshift, typically to 
an accuracy level of $10^{-3}$.  We discuss implications for figures 
of merit for dark energy science programs. 
\end{abstract} 

\maketitle

\section{Introduction \label{sec:intro}}

Understanding the nature of the dark energy accelerating the cosmic 
expansion is one of the premier questions in physics.  The answer 
offers the possibility of deep insights into the nature of spacetime 
and gravity, extra dimensions, the quantum vacuum, and possibly the 
unification of gravitation and quantum physics.  Precision mapping of 
the expansion history provides one path to characterizing the dark 
energy, in particular its equation of state and time variation. 

Guidance from theory is useful to predict observable signatures for 
cosmological probes such as distance and Hubble parameter measurements, 
in particular what level of accuracy is required to distinguish between 
models.  From a model one can predict distance-redshift relations etc.\ 
but the number of models is vast; one would like to identify model 
independent or at least generic characteristics of the dark energy. 
Indeed, such properties exist, as discussed in detail recently by 
\cite{cahndl}, for classes of behavior in the early time 
evolution of dark energy, valid for $z\gtrsim2$ when the dark energy 
does not strongly affect the background expansion. 

In this article we seek to extend characterization of the dark energy 
properties in terms of the equation of state to the entire 
observable history.  This requires a different approach, calibrating 
the evolution through a ``stretch'' relation between the amplitude of 
the time variation and the time variable or scale factor of the expansion. 
The calibration then provides a physical basis for a compact and highly 
accurate parametrization of the dark energy influence on observables. 

In \S\ref{sec:dyn} we examine several diverse models, looking 
for similarities and distinctions.  We introduce the calibration in 
\S\ref{sec:stretch} and discuss its relation to a standard parametrization 
of the equation of state.  \S\ref{sec:obs} examines the utility of the 
description and shows that it achieves robustness and accuracy at the 
$10^{-3}$ level, sufficient for next generation data.  We discuss some 
implications for figures of merit of dark energy science programs in 
\S\ref{sec:fom}. 
Those readers wanting to get right to the results could start in the 
middle of \S\ref{sec:stretch}.

\section{Dark Energy Dynamics \label{sec:dyn}} 

By examining the behavior of a diversity of dark energy models 
representing different physical origins, we can explore common and 
distinct elements within 1) a model as the parameters vary, 2) a 
family of models with some related property, and 3) different classes 
of models.  Families of models might consist of those with similar 
functional forms, e.g.\ polynomial potentials, while classes might 
be those with similar early time behaviors, e.g.\ thawing models or 
freezing models \cite{caldlin}. 

We choose five representative families ranging over different physics and 
different evolutionary histories.  These are the pseudo-Nambu Goldstone 
boson (PNGB) model, or cosine potential, that thaws and moves away from 
an early cosmological constant state $w=-1$, the family of polynomial 
potentials, also thawing, the supergravity-inspired SUGRA model that has 
early tracking behavior and then moves toward the cosmological constant 
state (freezing behavior), the modified gravity model of DGP braneworld 
cosmology and its family of $H^\alpha$ modifications of the Friedmann 
equation, also with freezing behavior, and the Albrecht-Skordis or 
exponential times polynomial potential, whose history cannot be classified 
as purely thawing or freezing. 

The dynamics is conveniently represented by the equation of state, or 
effective pressure to density ratio, $w$, and its variation $w'\equiv 
dw/d\ln a=\dot w/H$ where $a$ is the expansion or scale factor.  The 
Hubble parameter, or expansion rate, $H=\dot a/a$.  We work in units 
where $8\pi G=1$.

\subsection{PNGB Model \label{sec:pngb}} 

Protected from radiative corrections by a shift symmetry, this model 
possesses technical naturalness and is characterized by a symmetry 
energy scale $f$ \cite{friewaga}.  The potential reads 
\beq 
V(\phi)=V_\star\,[1+\cos(\phi/f)]\,, \label{eq:vpngb} 
\eeq 
with $V_\star$ setting the overall magnitude, hence related to the present 
dark energy density.  The equation of state, and the dynamics in general, 
is governed by $f$ and the initial field position $\phi_i$.  (It is 
convenient, as seen from the form of Eq.~(\ref{eq:vpngb}), to use 
$\phi_i/f$ instead of $\phi_i$.) 

\begin{figure}[!htb]
\begin{center}
\psfig{file=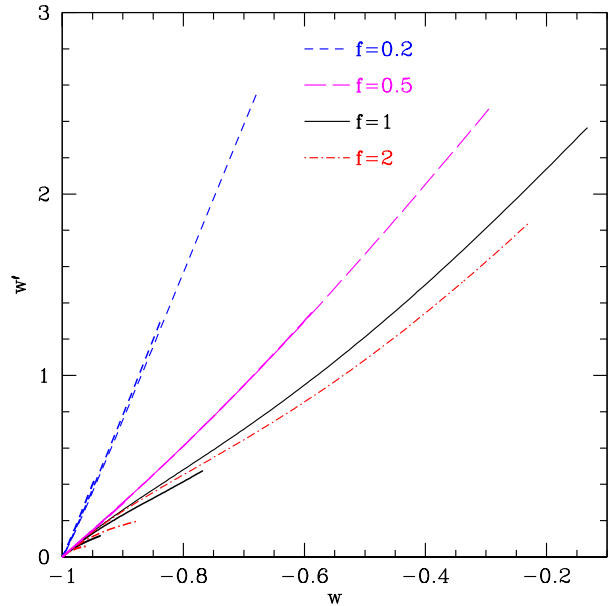,width=3.4in}
\caption{PNGB models fan out through phase space as their parameters 
vary (though still mostly within the thawing region).  At early times 
the models all start frozen at $(w,w')=(-1,0)$ and thaw, with the scale 
factor increasing along each curve, although at different rates in each 
case.  Here we end the tracks when $\op=0.72$.  
}
\label{fig:pngbfan}
\end{center}
\end{figure}

One can scan over the parameter space of these three variables and 
examine the evolutionary behavior and viability as a dark energy model. 
Figure~\ref{fig:pngbfan} shows a selection of trajectories in the 
$w$-$w'$ plane.  The time coordinate runs along these tracks, and can 
be thought of as the scale factor $a$ or the dark energy density fraction 
of the total energy density, $\omw(a)$.  
As we change $V_\star$ or $\omw=1-\om$, where $\om$ is the dimensionless 
present matter density, different points along 
a track for given $f$ and $\phi_i/f$ correspond to the present.  
In fact, for some parameter values the dark energy 
never dominates and the density is restricted to $\omw(a)< 
\Omega_{w,{\rm max}}<1$.  One can show that for fixed $\phi_i/f$, then 
$\omwmx\propto f^2$, so models with symmetry energy scales much less 
than the Planck energy, $f\ll1$, tend not to be 
viable. 

\begin{figure}[!htb]
\begin{center}
\psfig{file=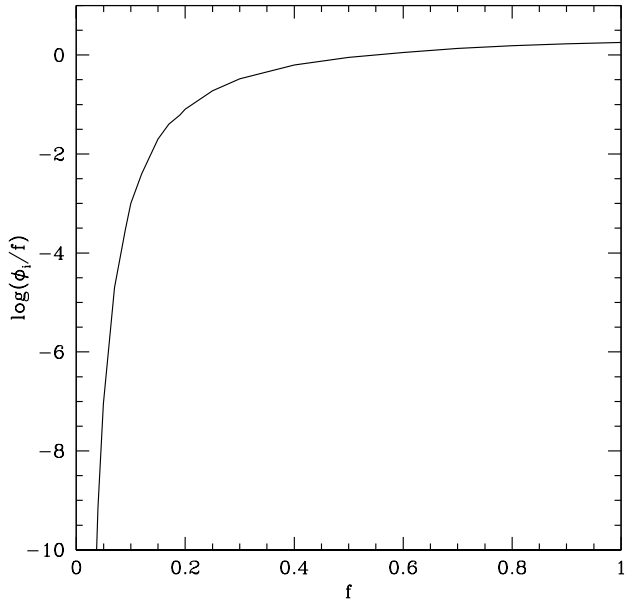,width=3.4in}
\caption{To achieve dark energy domination in the PNGB model before 
the field relaxes to its minimum, the initial field value $\phi_i$ must 
be small enough to give a long period of cosmological constant-like, or 
frozen, behavior.  For very steep potentials, i.e.\ low symmetry energy 
breaking scales $f$, the field must initially be extremely finely 
balanced near the top of the potential, with the curve showing the 
maximum $\phi_i/f$ allowed to achieve $\op\ge0.72$ at some point in the 
evolution. 
}
\label{fig:phif}
\end{center}
\end{figure}

Figure~\ref{fig:pngbfan} shows a wide selection of trajectories that 
reach $\omw=0.72$ at the present.  They fan out across the phase space, 
including ones that lie outside the conventional thawing region 
$3(1+w)>w'>1+w$ (although these start along $w'=3(1+w)$ at early 
times) \cite{caldlin}.  The exceptions have $f\ll1$, and are not 
generic in that 
for $f\ll1$ we must fine tune ever more strictly the initial 
condition $\phi_i/f$ in order to achieve such a present density.  
Figure~\ref{fig:phif} plots the allowed values of $\phi_i/f$, which 
decrease rapidly, 
roughly as $(\phi_i/f)_{\rm max}\sim e^{-1/f}$.  For example, when 
$f=0.1$, then the field must start exquisitely close to the top of 
the potential: rather 
than $\phi_i/f$ ranging freely over $[0,\pi]$, it is restricted to be 
less than $10^{-3}$.  For $f=0.05$, this becomes $\phi_i/f<10^{-7}$. 
Apart from unnaturalness, such values may run into physical problems 
such as a tachyonic instability \cite{spinodal,holman,kaloper}. 

In the future, the scalar field reaches the minimum of the potential 
and oscillates around it, giving an equation of state $w=0$ (matter-like) 
when averaged over many oscillations.  We discuss this further in 
comparison with the next model.

\subsection{Linear Potential \label{sec:linpot}} 

The linear potential tilts a flat potential, so the field rolls -- 
although it is frozen by the large Hubble friction at early times. 
The potential is given by 
\beq 
V(\phi)=V_i+(\phi-\phi_i)\,V'\,, \label{eq:vlp} 
\eeq 
where $V'$ is the slope parameter, a constant \cite{linde,wbg}.  
If the slope becomes too steep then the field never has time 
in its evolution to build up to appreciable energy density 
before the kinetic energy becomes substantial and 
$w>0$, causing the fractional energy density relative to the matter 
density to decrease with scale factor.  
The evolutionary tracks for this model fan out in the phase space 
within the thawing region (some examples for this and other models 
appear in \S\ref{sec:cross} in Fig.~\ref{fig:allfam}). 

\begin{figure}[!htb]
\begin{center}
\psfig{file=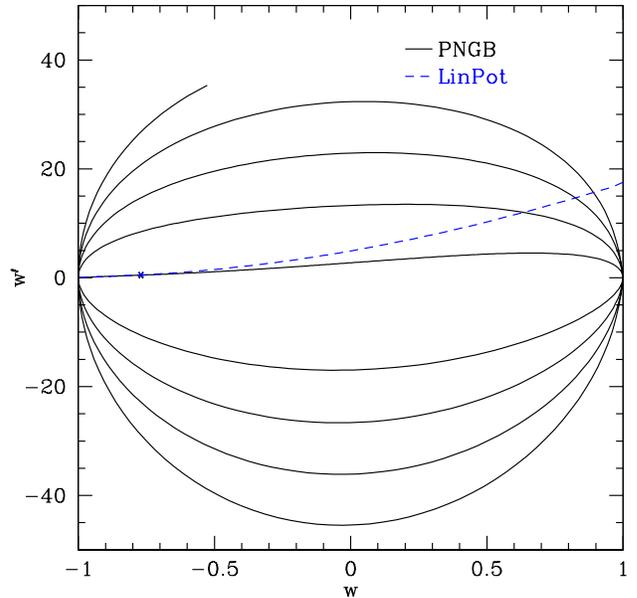,width=3.4in}
\caption{Long term evolutions of PNGB and linear potential models have 
distinct implications for the fate of the universe, both different from 
the cosmological constant case.  The PNGB field oscillates while the 
linear potential rolls to negative infinity.  Today the models shown 
have $w_0=-0.77$, with $w'_0=$0.47, 0.52 respectively. 
At $a=2$, the PNGB model is still on the innermost track, with $w=0.11$ 
(the curve end is at $a=6.6$), 
while the linear potential is off the plot, on the way to collapse. 
}
\label{fig:longfate}
\end{center}
\end{figure}

Figure~\ref{fig:longfate} shows the long time evolution of the PNGB 
vs.\ linear potential models, showing the similarity of the tracks at 
first, then the dramatic difference in the fate of the universe as 
the PNGB field oscillates, acting like matter in a time averaged sense, 
and the linear potential field shoots away, leading to deceleration 
and a cosmic doomsday collapse.

We also consider the related family of polynomial potentials, 
$V\sim\phi^n$ with $n=2$, 4.  These are also thawing models although 
their future behavior asymptotes to oscillation about a zero potential 
minimum.  Hence they do not runaway to negative potential and a rapid, 
doomsday collapse.  The equation of state during the oscillatory phase 
time averages to $w=(n-2)/(n+2)$ \cite{turner83}.

\subsection{SUGRA Model \label{sec:sugra}}

Tracking models have an early time attractor behavior that allows a 
large variety of initial conditions to give the same evolution in the 
matter dominated era, ameliorating fine tuning of initial conditions 
\cite{zws}.  
One example is the family of inverse power law potentials 
\cite{ratrapeebles}.  Including Planck scale corrections motivated by 
supergravity theory changes the potential to \cite{braxmartin} 
\beq 
V(\phi)=V_\star\,\phi^{-n} e^{\phi^2/2}. \label{eq:vsugra} 
\eeq 
This has a local nonzero minimum, or cosmological constant. 
The equation of state behavior is governed by the power law index $n$. 

Although the exponential factor has no effect on the attractor phase, 
it does permit a more rapid evolution after the field leaves that 
trajectory, moving the equation of state closer to $w=-1$.  Since 
the dark energy has $w=-2/(2+n)$ while on the attractor trajectory, 
one requires $n\ll1$ in the inverse power law model to accord with 
observations; this is somewhat eased for the SUGRA model.  
Today the field properties can cover a wide swath within the freezing 
region $0.2w(1+w)<w'<3w(1+w)$.

\subsection{Braneworld Gravity Model \label{sec:bw}} 

Even dark energy theories that do not involve scalar fields can be 
viewed in terms of effective dynamics, where the equation of state 
is defined in terms of the Hubble parameter $H(a)$ and its modified 
Friedmann equation: 
\beq 
w_{\rm eff}=-1-\frac{1}{3}\frac{d\ln\delta H^2}{d\ln a}, 
\eeq 
where $\delta H^2=H^2/H_0^2-\om a^{-3}$.  One example involving very 
different physics from scalar fields is the extension of gravity theory 
through extra dimensions.  This can lead to a modified Friedmann equation 
and effective equation of state \cite{DvaliTurn03,freese03} 
\beqa 
H^2&=&\rho_m(a)/3+(1-\Omega_m)\, H_0^2\, (H/H_0)^{\alpha}, \\ 
w&=& -\left[1 + \frac{\alpha}{2-\alpha} \,\Omega_m(a)\right]^{-1}, 
\eeqa 
where $\rho_m$ is the physical matter density and $\alpha$ is a 
parameter depending on boundary conditions between our four dimensional 
universe and the higher dimensional bulk volume.  The best motivated 
model in this family is DGP braneworld gravity, corresponding to 
$\alpha=1$ \cite{dgp,ddg}.  At early times the effective potential 
looks like an 
inverse power law \cite{cahndl}, with index $n=2\alpha/(2-\alpha)$, and 
so has tracking behavior.  At late times the field rolls asymptotically 
to a halt at a finite value of both the field and potential, acting as 
a cosmological constant.  Indeed the trajectories lie within the 
freezing region.

\subsection{Albrecht-Skordis Model \label{sec:as}} 

A scalar field potential with greater complexity  
is the Albrecht-Skordis \cite{asorig}, or exponential with polynomial 
prefactor, potential, motivated by string theory.  This has the form 
\beq 
V(\psi)=V_0\, [\chi(\psi-\beta)^2+\delta]\,e^{-\lambda\psi}, 
\eeq 
in the notation of \cite{barnardas}, with a more compact but 
equivalent notation being  
\beq 
V(\phi)=V_\star\,(1+A\phi^2)\,e^{-\lambda\phi}, 
\eeq 
where we shift to $\phi=\psi-\beta$, showing that only three 
parameters enter: $V_\star$, related to the dark energy density today, 
$\lambda$, and $A$.  Away from $\phi\approx 0$ this behaves like 
an exponential potential, a classic 
tracker, so the initial conditions are not very important 
\cite{wett88}.  Near 
$\phi=0$, the potential has a false minimum, so a field rolling 
through this region can have complicated dynamics, and indeed be 
trapped and oscillate about the nonzero potential minimum, eventually 
relaxing to a cosmological constant. 

\begin{figure}[!htb]
\begin{center}
\psfig{file=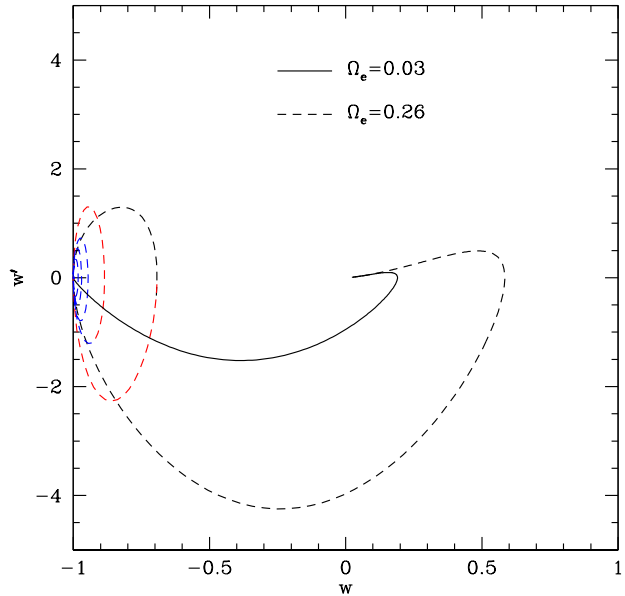,width=3.4in}
\caption{Albrecht-Skordis model acts like a tracer at early times, 
with a constant energy density fraction $\Omega_e$ and $w=0$ in the 
matter dominated era, before oscillating around the 
nonzero minimum of the potential.  We change the line thicknesses (and 
colors) at $z=2$, 1, and 0.  The oscillations are invisible for more 
viable $\Omega_e=0.03$ case. 
}
\label{fig:astrack}
\end{center}
\end{figure}

\begin{figure}[!htb]
\begin{center}
\psfig{file=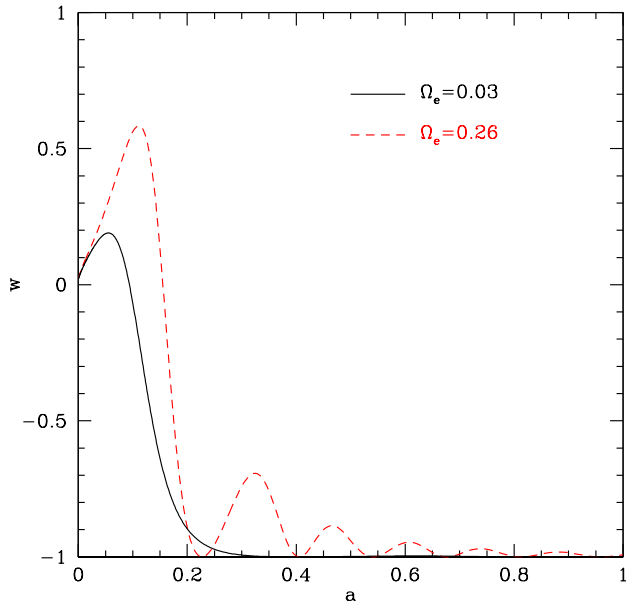,width=3.4in}
\caption{Equation of state $w(a)$ shows an alternate view 
of the evolution in Fig.~\ref{fig:astrack}.  Note that for model parameters 
that do not violate early matter domination, the behavior relaxes swiftly 
to a cosmological constant, as shown by the solid, black curve. 
}
\label{fig:aswa}
\end{center}
\end{figure}

Figure~\ref{fig:astrack} show trajectories for different parameter 
values, illustrating the wide variety of possible behaviors.  In 
addition, Fig.~\ref{fig:aswa} plots the equation of state $w(a)$ so 
one has another view of the damped, oscillatory evolution.  Note 
that while the field sees an exponential potential, away from the 
false minimum, it exhibits not only tracking but tracing behavior -- 
the dark energy equation of state is equal to the background, e.g.\ 
matter dominated, equation of state $w_b$.  This means that the dark 
energy density is then a constant fraction of the background density, 
given by $\Omega_{w,{\rm trace}}=3(1+w_b)/\lambda^2$ 
\cite{ferrjoyce,copelw}.  
So as not to violate primordial nucleosynthesis or cosmic  
microwave background constraints, this 
requires the contribution to be no more than a few percent.  We show 
the dynamics for two cases, the first using the parameter values in 
\cite{barnardas}, corresponding to $\lambda=3.4$ and $A=106.7$, which 
has an early dark energy fraction $\Omega_e=0.26$ during matter 
domination ($\Omega_e=0.35$ during radiation domination), and the 
second using $\lambda=10$, keeping $A$ the same, giving $\Omega_e=0.03$ 
(0.04) during matter (radiation) domination, close to the upper limit 
allowed \cite{doranlim,beanlim}.

\begin{figure}[!htb]
\begin{center}
\psfig{file=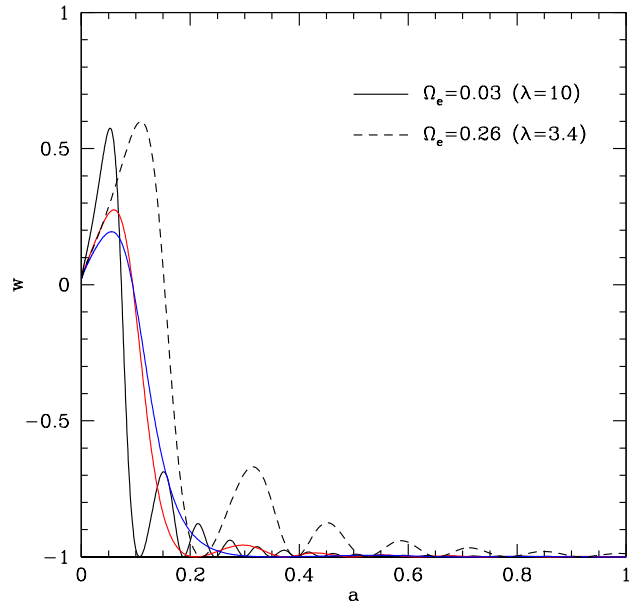,width=3.4in}
\caption{The amplitude of oscillations in the equation of state is 
governed by $\lambda^2/A$ and the period goes as $\lambda^{-1}$ (for fixed 
$\lambda^2/A$).  The figure shows the equation of state history for 
$\lambda = 10$ ($\Omega_e=0.03$ during matter domination) as solid 
curves, for $\lambda^2/A=0.1$, 0.5, 0.9, from highest peak to lowest, 
and for $\lambda=3.4$ ($\Omega_e=0.26$), with $\lambda^2/A=0.1$, as a 
dashed curve.  
For appreciable oscillations $\lambda^2/A$ must approach zero, but for 
allowed (large) values of $\lambda$ any oscillations damp away for 
$z\lesssim3$. 
}
\label{fig:asosc}
\end{center}
\end{figure}

For values of $\lambda$ allowed by nucleosynthesis and CMB limits, 
$\lambda\ge10$, the oscillations are absent or negligible.  One can 
show that the amplitude of the oscillations depends predominantly on 
the ratio $\lambda^2/A$ (e.g.\ define $\varphi=\lambda\phi$ and the 
potential only explicitly contains the parameter combination 
$\lambda^2/A$).  If this combination exceeds one, then there is no 
minimum but merely a slight local lessening of the exponential slope, 
and hence no oscillations.  The amplitude increases as $\lambda^2/A$ 
approaches zero.  However, since $\lambda\ge10$, small values of 
$\lambda^2/A$ require $A\gtrsim1000$, seemingly unnatural.  
Furthermore, the 
period of the oscillations is given by the effective mass and is 
inversely proportional to $\lambda$ for fixed $\lambda^2/A$, and so 
for allowed $\lambda$ the oscillations will be negligible for 
$z\lesssim3$.  These behaviors are illustrated in Fig.~\ref{fig:asosc}.

\subsection{Cross Comparison \label{sec:cross}} 

\begin{figure}[!htb]
\begin{center}
\psfig{file=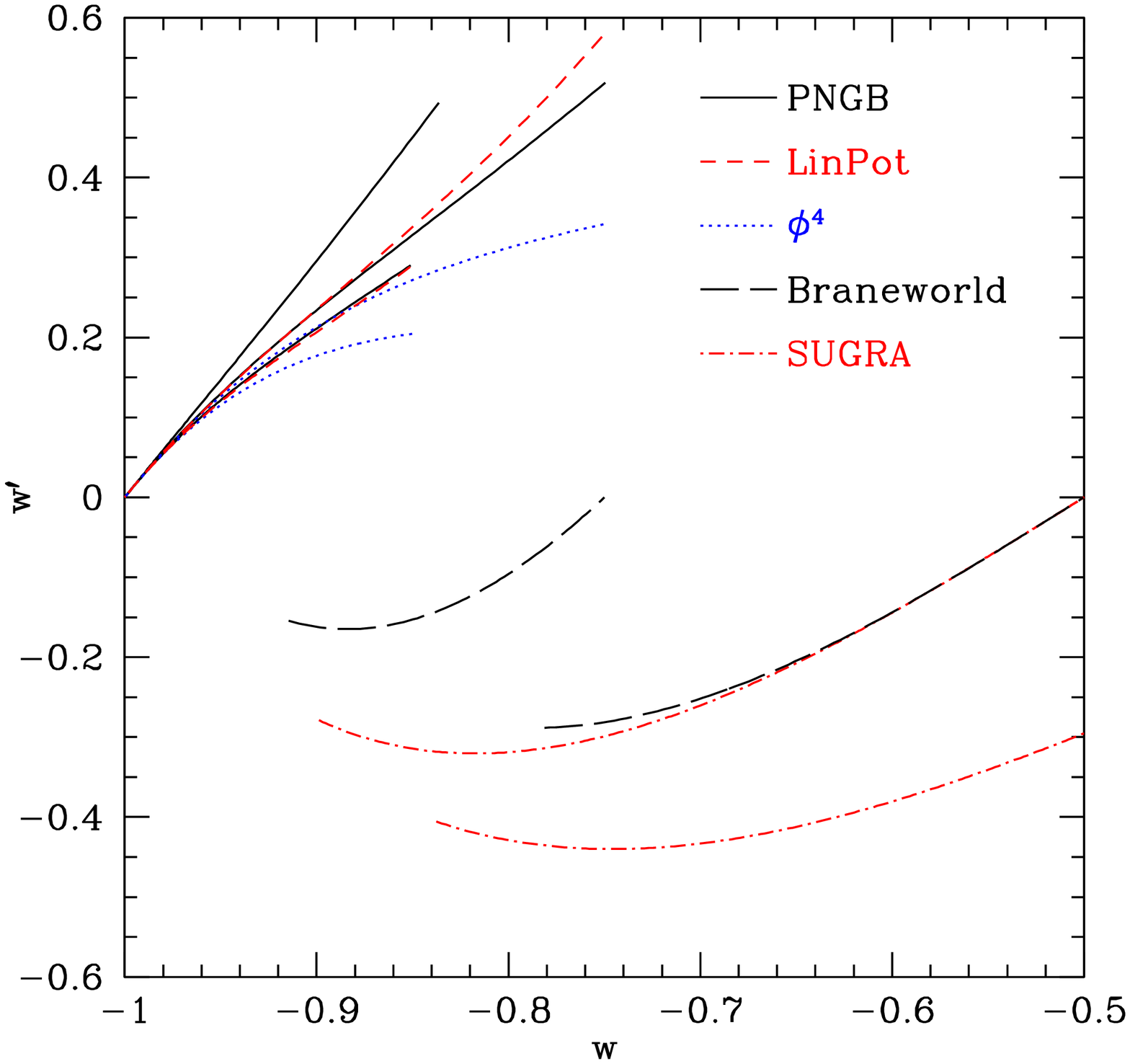,width=3.4in}
\caption{Representative models considered in this section are plotted 
for various parameter values in the $w$-$w'$ phase space.  Solid, black 
curves are PNGB, short-dashed, red curves are for the linear potential, 
dotted, blue curves are for $\phi^4$, long-dashed, black curves for 
the braneworld model ($\alpha=1$ DGP and $\alpha=0.5$), and dot-dashed, 
red curves for SUGRA. 
}
\label{fig:allfam}
\end{center}
\end{figure}

To compare the behaviors of different families, we plot selected 
representatives in Fig.~\ref{fig:allfam}.  Varying the parameters 
within each model, as well as considering different models, spreads 
the evolution over regions of the $w$-$w'$ phase space.  
Generally we see both similarities and distinctions between models and 
between families.  One must also take into account the time coordinate 
along the curves, so that crossing of trajectories does not mean 
they have identical properties at any one moment.  
We plot the trajectories up to when the dark energy density is $\op=0.72$.

We could extend the curves into the future, as was done in 
Figs.~\ref{fig:longfate} and \ref{fig:astrack}.  The $\phi^4$ potential 
will eventually settle at $w=1/3$, acting as radiation, as discussed 
in \S\ref{sec:linpot}, after oscillating around the minimum.  
Note that the SUGRA, DGP/$H^\alpha$, and Albrecht-Skordis models all 
have nonzero minima, i.e.\ hidden cosmological constants, so they 
settle to $w=-1$.  The SUGRA field does not oscillate around the minimum 
because it approaches it with low kinetic energy, freezing to the 
cosmological constant state;  the $H^\alpha$ family has only an 
asymptotic minimum, also approached by freezing.  

As an alternative to showing 
the evolution of varied models at all times, we can take a slice 
at a particularly time, say when $\op=0.72$, and construct phase space 
curves where the parameters of a potential vary along the curve. 
This can clear the illusion of overlap in behavior and provides an 
intermediate step toward the calibration in the next section. 

\begin{figure}[!htb]
\begin{center}
\psfig{file=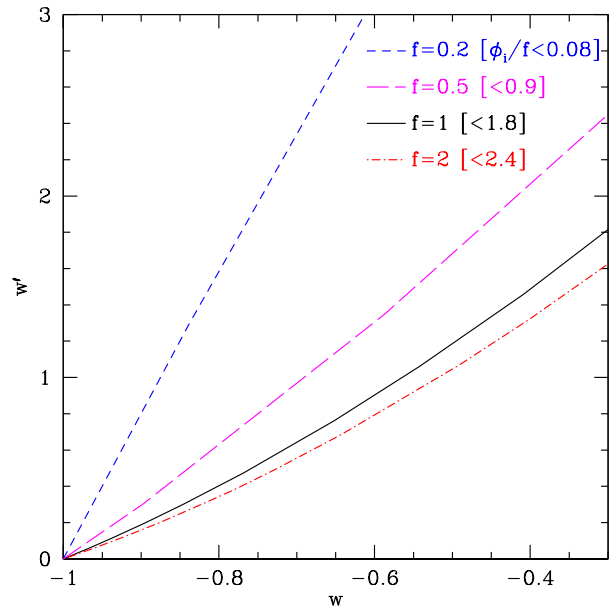,width=3.4in}
\caption{These curves for the PNGB model correspond to a scan over the 
potential parameter space to find those values where $\op=0.72$.  Each 
curve is for a different energy scale $f$, with the parameter $\phi_i/f$ 
running along each curve, from zero at $w=-1$ to a maximum possible value 
shown in brackets. 
}
\label{fig:pngbcurve}
\end{center}
\end{figure}

Figure~\ref{fig:pngbcurve} gives an example of this for the PNGB model, 
where the parameter running along the curve is the initial field position 
$\phi_i/f$.  That is, every point along any curve has $\op=0.72$ today, 
but corresponds to a different set of parameters for the potential and 
a different evolutionary behavior.  This illustrates that different 
symmetry energy scales $f$ define 
distinct paths to achieving a given dark energy density.  (Of course 
not all of these are viable, with large values of $\phi_i/f$ along 
each curve corresponding to $w$ far from $-1$, and small values of $f$ 
suffering from the extreme fine tuning problem discussed in 
\S\ref{sec:pngb}.)   However, by evaluating the equation of state and 
its time variation at a single time, we lose all dynamical information. 
In the next section we combine the advantages of the parameter scan 
with those of the evolutionary trajectories.

\section{Stretching Dark Energy \label{sec:stretch}} 

To keep the dynamics central, we want to preserve in some way the 
temporal information, i.e.\ the field evolving from its high redshift 
state along a trajectory describing the 
equation of state and its time variation.  However, we are free to 
rescale the time coordinate and define a time variation other than 
$w'=dw/d\ln a$.  In particular, we can ask whether there is a global 
transformation that in some way calibrates the dark energy characteristics. 
We call this the evolutionary stretch factor. 

Stretching the time variation by different amounts at different times 
effectively introduces additional evolution beyond the scalar field 
behavior, so we consider a constant stretch factor, a simple 
renormalization.  That is, we take $w'(a)\to w'(a)/a_\star$.  Now, since 
realistic observations cannot map out the detail of the equation of 
state function, we seek to condense the information on the evolution to 
a set that is robustly constrained by data.  Overcompression loses 
important physical properties while undercompression leads to 
uninformatively large uncertainties.  In the next section we will test 
the full stretch prescription to 
ensure that neither case occurs.  To begin with, consider evaluating our 
new time variation quantity at a particular scale factor; furthermore, 
to keep the number of parameters in the stretch prescription to a minimum, 
we choose this scale factor to be the same as the stretch factor $a_\star$. 
That is, the procedure can be viewed illustratively as 
\beq 
w'(a)\to \frac{w'(a)}{a_\star}\to \frac{w'(a_\star)}{a_\star}\,. 
\eeq 

For evaluating the value of the equation of state function itself, $w(a)$, 
we also avoid 
choosing an arbitrary scale factor.  This leaves us with two choices: 
either $a_\star$ or the present epoch, $a=1$.  If we choose $a_\star$, 
then this procedure merely chooses a single point along the evolutionary 
trajectory, losing much of the global information.  Thus we adopt 
$w_0=w(a=1)$ and examine the dark energy characteristics in the plane 
of the two parameters, $w_0$ and $w'(a_\star)/a_\star$ ($=dw/da(a_\star)$), 
to see if there is 
indeed a normalizing relation for the time evolution. 

\begin{figure}[!htb]
\begin{center}
\psfig{file=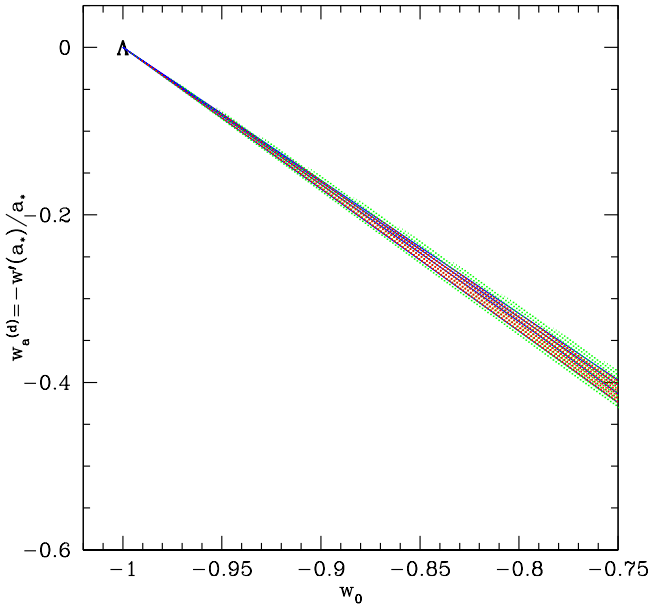,width=3.4in}
\caption{Defining a new time variation variable $w_a^{\rm(d)}$ from 
$w'$ calibrates the PNGB model into a tight locus; compare the spread 
in Fig.~\ref{fig:pngbcurve}.  Solid lines are for fixed $f$ parameter, 
the shading shows the range of behaviors for $f\in[0.2,5]$.  The lighter 
shading shows the effect of also scanning over $\om=0.25$--0.31. 
}
\label{fig:pngbstretch}
\end{center}
\end{figure}

Figure~\ref{fig:pngbstretch} shows clearly that this prescription 
calibrates the evolution of the PNGB model.  Instead of the fan of 
trajectories spreading through the $w$-$w'$ phase space, as in 
Fig.~\ref{fig:pngbcurve}, we now have 
a tightly calibrated, one parameter relation in the $w_0$ vs.\ 
$w'(a_\star)/a_\star$ plane.  
Despite scanning over the model space of $f$ and $\phi_i/f$, this stripe 
is narrow and well defined.  Points within the stripe represent individual 
realizations of the PNGB model with choices of the symmetry energy scale 
ranging over the physically reasonable range $f\in[0.2,5]$ and initial 
field position covering from 0 to the maximum value that allows 
$\op\approx0.7$.

This tight calibration spreads little if we vary the present dark energy 
density as well as the potential parameters themselves.  Allowing $\op$ 
today to range over 0.69-0.75 gives the slightly wider, lightly shaded region. 

\begin{figure}[!htb]
\begin{center}
\psfig{file=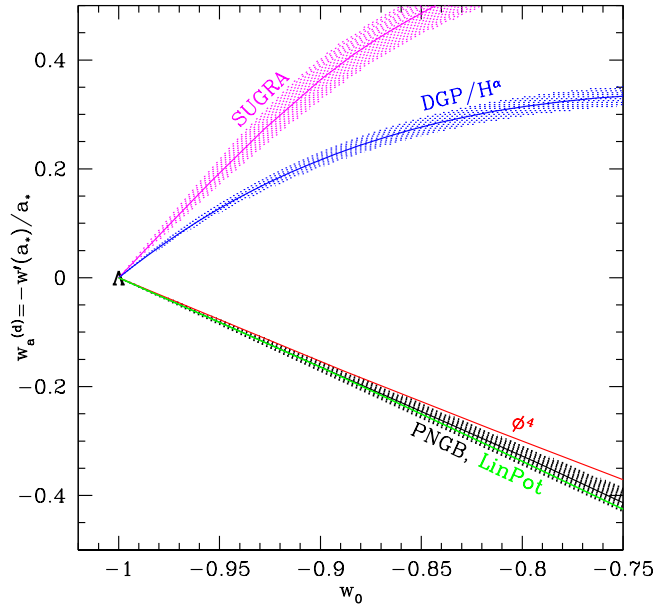,width=3.4in}
\caption{In terms of the calibrated dark energy parameters $w_0$ and 
$w_a^{\rm (d)}$, models and families lie in tightly homogeneous regions, 
in comparison to Fig.~\ref{fig:allfam}, showing the same models before 
calibration.  We here vary over all parameters in the potentials.  
Shading shows the effect of scanning over $\pm0.03$ in $\om$ (we omit 
the shading for $\phi^4$ and linear potential models to minimize confusion; 
the width would be about half that shown for PNGB).  Thawing models, 
despite their differences in $w$-$w'$, are nearly identical once 
calibrated.  Distinctions from freezing models, and between freezing 
models, become highlighted with calibration. 
}
\label{fig:manystretch}
\end{center}
\end{figure}

Calibration succeeds for the other dark energy models considered as well, 
covering 
a wide range of physical origins.  Indeed, all the thawing models are 
closely related, nearly forming a single family under the calibration. 
The similarities extend to defining a single stretch parameter $\as=0.8$ 
for the entire thawing class.  Freezing fields also can be calibrated, 
with a uniform stretch parameter $\as=0.85$, though the families stay 
more distinct within the freezing class.  Figure~\ref{fig:manystretch} 
shows the tight relations of the different dark energy models, in strong 
contrast with the ``fan'' nature of Fig.~\ref{fig:allfam}.

From the form of the stretch calibrated time variation, $w'(\as)/\as$, 
we can recognize this as nearly identical to $w_a$, the dark energy 
variable in standard use, defined by \cite{linprl} as $w_a=-w'(\as=0.5)/0.5$ 
to fit the equation of state function by $w(a)=w_0+w_a^{\rm(w)}(1-a)$.  
The superscript w indicates that the value of $\as$ was chosen to fit 
$w(a)$.  Here, however, we defined the equivalent of $w_a$ to calibrate 
dark energy families.  This resulted in $\as=0.8$ for the thawing class 
and $\as=0.85$ for the freezing class.  An interesting further implication 
is that the ``new'' form 
\beq 
w(a)=w_0-\frac{w'(\as)}{\as}(1-a)=w_0+w_a^{\rm(d)}(1-a) \label{eq:wad} 
\eeq 
has excellent accuracy when fitting the observables of distance and 
Hubble parameter, as we discuss next.

\section{Observing Dark Energy \label{sec:obs}} 

While the form (\ref{eq:wad}) was just shown useful in interpretation 
of dark energy theory, we should also investigate its utility for 
interpreting dark energy observations.  
Three related, but slightly different, physical bases exist 
for using the form $w(a)=w_0+w_a(1-a)$ to characterize dark energy: 
this can be interpreted as 1) a fitting formula to the equation of 
state, 2) a calibration relation for families of dynamics in the 
$w$-$w'$ plane, or 3) a fitting formula for observables such as 
distances and the Hubble expansion rate.  The last two in particular 
are closely related and give similar results; indeed, when models do 
not deviate greatly from cosmological constant behavior the two 
approaches are almost identical.  

We now explore the accuracy of the form (\ref{eq:wad}) in fitting the 
exact distance-redshift and Hubble parameter-redshift relations for 
the diverse dark energy models discussed in \S\ref{sec:dyn}.  

Table~\ref{tab:waacc} summarizes the accuracies on $d$ and $H$ for 
a diverse range of models.  These are generally good to the 
$10^{-3}$ level.  Models closer to $\Lambda$ would have better 
fits than shown here; models further from $\Lambda$ are not favored by 
current data.  For simplicity we henceforth denote the calibrated fit 
parameter simply as $w_a$.

\begin{table}[htbp]
\begin{center}
\begin{tabular*}{0.9\columnwidth} 
{@{\extracolsep{\fill}} l c c }
\hline
Model & $\delta d/d$ & $\delta H/H$ \\ 
\hline
PNGB ($w_0=-0.85$) & 0.05\% & 0.1\% \\ 
PNGB ($w_0=-0.75$) & 0.1\% & 0.2\% \\ 
Linear Pot.\ ($w_0=-0.85$) & 0.05\% & 0.1\% \\ 
Linear Pot.\ ($w_0=-0.75$) & 0.1\% & 0.3\% \\ 
$\phi^4$ ($w_0=-0.85$) & 0.01\% & 0.04\% \\ 
$\phi^4$ ($w_0=-0.75$) & 0.02\% & 0.06\% \\ 
Braneworld ($w_0=-0.78$) & 0.03\% & 0.07\% \\ 
SUGRA ($n=2$) & 0.1\% & 0.3\% \\ 
\hline 
SUGRA ($n=11$) & 0.1\% & 0.3\% \\ 
Albrecht-Skordis ($\Omega_e=0.03$) & 0.01\% & 0.02\% \\ 
Albrecht-Skordis ($\Omega_e=0.26$) & 0.1\% & 0.4\% \\ 
\hline 
\end{tabular*}
\caption{Accuracy of $w_0$-$w_a$ in fitting the exact distances and 
Hubble parameters for various dark energy models.  These numbers 
represent global fits over all redshifts (except for the last three cases, 
where the fit covers $z=0$-3, due to early dark energy: 
see \S\ref{sec:fom}).  Better fits can be found over finite redshift 
ranges. } 
\label{tab:waacc}
\end{center}
\end{table}

We could push the accuracy even further by minimizing the deviation 
not globally, over the entire range $a\in[0,1]$, but over a particular 
epoch, say $a\in[0.5,1]$.  However, we retain the global fit in general. 
Also, we have not taken advantage of the degree of freedom of $w_0$, 
which could improve the fits. 
We emphasize that the stretch factor is a function of the dark energy 
physics and not dependent on the experiment, priors, etc.\ (in 
distinction from a pivot redshift or pivot equation of state value). 

Note that the results from this prescription also answer the important 
question of whether the calibration 
procedure preserves the information faithfully to the precision level 
of the data, or over- or under-compresses the model characteristics. 
A one parameter approach such as a constant value of $w$ would have 
errors of order 1--2\% in distance and up to 3\% in Hubble parameter 
for the models we considered.  This is insufficient for forthcoming 
observations.  Conversely, since the two calibrated parameters 
of $w_0$ and $w_a$ 
map the observables to better accuracy than expected from next generation 
data, these two parameters suffice and the data precision does not call 
for further equation of state parameters. 

This is not to say that 
some models could not exist where a third parameter carries information, 
but such models may not be generic or natural; the wide range of models 
considered here has no use for one.  If we reach the stage of probing 
the cosmic expansion history below the $10^{-3}$ precision level, we 
should revisit the question of a further calibration parameter. 

Finally, this prescription is meant to help us find our way through 
the dark forest \cite{dante} of models of cosmic acceleration, making 
accurate, more or less model independent assessments.  Once precision 
data exist, they should be analyzed for every model of interest 
and within every applicable fit technique, parametric and nonparametric.  
We have seen that until we reach that point $w_0$-$w_a$ serves as a 
robust indicator and guide for predicting and comparing cosmological 
probe information.

\section{Figures of Merit \label{sec:fom}} 

The accuracy of the $w_0$ and $w_a$ form, defined in the manner discussed 
here, for characterization of observable properties of dark energy 
is at a level of order $10^{-3}$, sufficient for next generation 
experiments.  
The calibration into tight families of equation of state properties, 
as seen in Fig.~\ref{fig:allfam}, suggests that not all combinations 
of $w_0$ and $w_a$ are of equal insight.  For example, one might 
distinguish models in the thawing class from the cosmological constant 
and from each other by constraining the combination varying exactly 
along the calibrated curve. 

Since this curve is nearly straight, we can characterize it by slope 
$m_t$, and define a new variable 
\beq 
w^t_\parallel=w_0+m_t w_a\,,
\eeq 
where the derivative with respect to this parameter runs parallel to 
the calibrated curve.  Hence, determining 
$w^t_\parallel$ localizes the behavior and distinguishes the specific 
dark energy characteristics.  
The narrowness of the calibrated region means that it is not so useful 
within thawing models to constrain the direction 
perpendicular to the curve.  

One can define a similar variable for the freezing class, although here 
the families are more spread out, so the slope is more of an average 
than a well defined value, 
\beq 
w^f_\parallel=w_0+m_f w_a\,. 
\eeq 
Values of $m_t=-1.75$ and $m_f=3.5$ are reasonable choices.  
Note that the combinations $w^f_\parallel$ and $w^t_\parallel$ 
are not orthogonal, so the variable defined for each 
class does have utility in constraining the other class as well.  
For example, along the PNGB curve of Fig.~\ref{fig:manystretch} the 
parameter $w^t_\parallel$ runs from $-1$ to 0, while $w^f_\parallel$ goes 
from $-1$ to $-2.2$; along the SUGRA curve $w^t_\parallel=-1$ to $-1.7$ 
while $w^f_\parallel=-1$ to $+0.9$.  This shows that each parameter, 
while optimized for a given physics question, does carry information 
on the other class. 

Thus, knowledge of either parameter $w^t_\parallel$ or 
$w^f_\parallel$ answers the key questions of distinction 
from a cosmological constant, distinction between models, and to an 
extent distinction between classes.  Constraining both parameters 
tightens the distinguishing ability, especially between classes, and 
provides a crucial crosscheck of the framework. 

It does not seem natural or effective to combine the uncertainties in 
estimating these variables from observations into a single number, e.g.\ 
$\sigma(w^t_\parallel)\times\sigma(w^f_\parallel)$, since they represent 
very different physics.  Moreover, further investigation is needed into the 
optimum values for $m_f$, $m_t$ and other issues before defining ultimate 
figures of merit, if this is even possible.  However, the tightness of 
the calibration does imply that some combinations of $w_0$ and $w_a$ 
will provide insight into the nature of dark energy.  Therefore, 
knowledge of the uncertainties  $\sigma(w_0)$ and $\sigma(w_a)$ and 
their covariance are the main ingredients for a variety of future 
figures of merit that might be developed. 

Finally, we note that the accuracy of the $w_0$-$w_a$ form does start 
to degrade to the $10^{-2}$ level as dark energy becomes increasingly 
important in the early universe around $z\gtrsim10^3$, upsetting 
standard matter domination.  See, for example, the 
last three models in Table~\ref{tab:waacc}, where the dark energy 
equations of state at recombination are $w\approx-0.15,$ 0, 0, 
respectively.  It could be useful to treat such early dark energy models 
as a separate class, and include constraint on the dark energy density 
$\Omega_e$ at recombination (which can best be done through growth 
probes) as another desideratum for a dark energy science program.

\section{Conclusions \label{sec:concl}} 

Having investigated a diverse group of dark energy models to explain 
the acceleration of the cosmic expansion, we find a homogeneous 
``stretch'' relation that calibrates the time variation behavior into 
tight families.  This stretch factor is closely related to the standard 
time variation measure $w_a$, and we verify that the equation of state 
form $w(a)=w_0+w_a(1-a)$, with $w_a$ now treated as a fit parameter to 
observables, delivers fractional accuracy at the $10^{-3}$ level. 

Such accuracy is sufficient for next generation data and the $w_0$-$w_a$ 
form can be viewed as an appropriate compression of the expansion 
history information that can be extracted from such observations.  That 
is, this form neither overcompresses (loses important information) nor 
undercompresses (lacks additional leverage).  This indicates there is 
no need nor generic benefit for going to a third parameter. 
Note that \cite{barnardchi} saw similar compression and tight relations 
within a principal component analysis relying on many modes. 

To gain insight into the nature of dark energy, particular combinations 
of $w_0$-$w_a$ may have enhanced leverage and hence merit, separating 
the cosmological constant from the thawing class, each from the 
freezing class, and possibly zeroing in on specific models within a 
class.  The calibration, and its robustness and accuracy in accounting 
for the observable relations, offers a well-defined method for 
assessing the next generation dark energy science program.  
Interpretation of those observations should offer promising insights 
into the physics of the accelerating universe.

\acknowledgments 

We thank Andy Albrecht and Michael Barnard for helpful discussions. 
This work has been supported in part by the Director, Office of Science, 
Office of High Energy Physics, of the U.S.\ Department of Energy under 
Contract No.\ DE-AC02-05CH11231. 


\end{document}